\documentclass{llncs}
\usepackage{enumitem}

\newcommand{\id}{\textrm{id}}
\newcommand{\toSet}{\textit{toSet}}
\newcommand{\val}{\textit{val}}
\newcommand{\ida}[1]{\textrm{id}(\texttt{#1})}

\newcommand{\implique}{\ensuremath{\Longrightarrow } }

\newcommand{\eqrel}{\ensuremath{\sim}}

\newcommand{\seq}[1]
{\begin{tabular}{l}
#1
\end{tabular}}

\newcommand{\while}[3]
{\begin{tabular}{l|ll}
    \multicolumn{3}{l}{
    
    \begin{tabular}{ll}
    \textbf{while}  & \begin{minipage}[t]{#1cm}#2\end{minipage}
    \end{tabular}

    }\\[-0.5em]
&&\seq{#3}\\[-0.5em]
&\multicolumn{2}{|l}{\ }\\
\cline{2-3}
\end{tabular}}

\newcommand{\myif}[3]
{\begin{tabular}{l|ll}
    \multicolumn{3}{l}{
    
    \begin{tabular}{ll}
    \textbf{if}  & \begin{minipage}[t]{#1cm}#2\end{minipage}
    \end{tabular}

    }\\[0.3em]
&&\seq{#3}\\[-0.5em]
&\multicolumn{2}{|l}{\ }\\
\cline{2-3}
\end{tabular}}

\title{Experimental Evaluation of\\ a Method to Simplify Expressions}

\author{Baudouin Le~Charlier}
\institute{Universit\'e catholique de Louvain, Belgium}

\begin{document}

\maketitle

\begin{abstract}
We present a method to simplify expressions in the context of an equational theory.
The basic ideas and concepts of the method have been presented previously elsewhere but here we tackle the difficult task of making it efficient in practice, in spite of its great generality. We first recall the notion of a collection of structures, which allows us to manipulate very large (possibly infinite) sets of terms as a whole, i.e., without enumerating their elements. Then we use this tool to construct algorithms to simplify expressions. We give various reasons why it is difficult to make these algorithms precise and efficient. We then propose a number of approches to solve the raised issues. Finally, and importantly, we provide a detailed experimental evaluation of the method and a comparison of several variants of it. Although the method is completely generic, we use (arbitrary, not only two-level) boolean expressions as the application field for these experiments because impressive simplifications can be obtained in spite of the hardness of the problem. 

\end{abstract}

\section{Introduction}
\label{sect:introduction}


Many people (engineers, logicians, mathematicians, students and experienced practitioners) are faced, in their everyday practice, with the task of simplifying expressions. This is useful or even necessary to understand the meaning of a formula resulting from a long computational effort, or to optimize the implementation of a compiler or the design of a logic circuit, for example. In this paper, we propose a new and automatable method to perform such simplifications based on a finite set of equations (or axioms) formalizing the meaning of the expressions to be simplified.
Since the method is general, namely parameterized by the chosen set of equations, it is applicable to many mathematical and logical domains.\footnote{Because of that, in this paper, we consider the words term, expression, and formula as synonyms: We prefer not to choose a single name and stick to it, because we want to be free to use the most accepted word in any application context.} The novelty and power of the presented method stem from the use of a powerful data structure, called \emph{collection of structures}, introduced in \cite{DBLP:conf/lpar/CharlierA15,DBLP:conf/sycss/CharlierA16} and thorougly studied in \cite{atindehouPhDThesis}.
Collections of structures allow us to represent very large sets of equivalent terms compactly. Axioms can be used to add new equivalent terms to the collection of structures without enumerating the terms. Simplification takes place when new, simpler terms appear in the collection of structures. 
Early attempts to using collections of structures to build simplification algorithms have been proposed in \cite{atindehouPhDThesis,DBLP:conf/lpar/CharlierA15}. The contributions of this paper are the following: (a) We give precise guidelines for constructing simplification algorithms that are more accurately defined and much more efficient than our previous proposals, (b) we explain in detail how and why simplification takes place, (c) we provide a thorough experimental evaluation of a large number of variants of a reference algorithm, using boolean expressions as our application field.

\section{Previous Work: Collections of Structures}
\label{sect:structures}

In this preliminary section, we summarize the main definitions, properties, and results about collections of structures. More explicit information and examples can be found in \cite{atindehouPhDThesis,DBLP:conf/sycss/CharlierA16}.

We assume given an (implicit) set of \emph{all terms}.  Conceptually, we only use \emph{binary terms}, of the form $f(t_1, t_2)$, and a unique, `dummy', constant \textit{null}. Regular constants such as $a$ and non binary terms such as $h(g(a, b, c)) $ can be simulated by binary terms and \textit{null} as done e.g., in \cite{Downey:1980:VCS:322217.322228} and \cite{NO:RTA-2005}.
In practice however, we stick to standard notation for writing examples. Let $T$ be a set of terms. We say that $T$ is \emph{sub-term complete} if every sub-term of a term of $T$ belongs to $T$. Assuming that $T$ is sub-term complete,
we say that a relation $\eqrel$, over $T$, is a \emph{congruence} if it is an equivalence relation such that   
$$
\left.\begin{array}{c}
s_i\ \eqrel\ t_i \:\:(i=1,2)\\
 f(s_1, s_2)\in T
\end{array}\right\}
\implique
f(s_1, s_2)\ \eqrel\ f(t_1, t_2)
$$
Note that $\eqrel$ is not a relation over all terms, in general, so that our definition is not standard.
To represent sets of terms, we use \emph{structures}. A structure is of the form $f(i_1, i_2):i$ where $f$ is a function symbol, and $i_1$, $i_2$, and $i$ are  \emph{identifiers of sets of structures}, or simply \emph{identifiers}, for short. We use natural numbers as identifiers and we choose a special identifier $i_\textit{\tiny null}$ to represent the term \textit{null}. We call $f(i_1, i_2)$, the \emph{key} of the structure. A \emph{collection of structures} is a finite set of structures. Let $E$ be a collection of structures. Let $I$ be the set of identifiers used by the structures of $E$. We write $E_i$ to denote  the set of structures of $E$ that are of the form $f(i_1, i_2):i$. Therefore, we have $E = \bigcup_{i\in I} E_i$.
By definition, $E$ \emph{denotes} the (unique) family of sets of terms $(T_i)_{i\in I}$ such that a term $f(t_1, t_2)$ belongs to $T_i$ whenever $E_i$ contains the structure $f(i_1, i_2):i$ and $t_1$, $t_2$ belong to $T_{i_1}$ and $T_{i_2}$, respectively. 
We only use \emph{well-formed} collections of structures (see \cite{DBLP:conf/sycss/CharlierA16}), which  fulfill a simple condition that ensures that no set $T_i$ is empty. A well-formed collection of structures is \emph{normalized} if it does not contain two different structures  $f(i_1, i_2): i$ and $f(i_1, i_2): i'$ (with the same key but different set identifiers).  It can be proved \cite{DBLP:conf/sycss/CharlierA16} that $E$ is normalized if and only if $\{T_i\ \mid \: i\in I\}$ is a partition of $T = \bigcup_{i\in I} T_i$ and $T_i \neq T_j$ ($\forall i,j\in I: i\neq j$).  In that case, the sets $T_i$ are the equivalence classes of a congruence $\eqrel$, as defined above, and we say that $\eqrel$ is the \emph{abstract denotation} of $E$. In the following, unless stated, we assume that $E$ is normalized.

There are four main operations to handle structures: \textit{toSet}, \textit{substitute}, \textit{normalize} and \textit{unify}. 
The operation \textit{toSet} adds a term $f(t_1, t_2)$ to a collection of structures $E$. It returns the identifier $i$ of the set of structures $E_i$ to which the term belongs. (More exactly, the term belongs to the set $T_i$ denoted by $E_i$. For simplicity, we often use this shortcut.) It first recursively computes the identifiers $i_1$ and $i_2$ corresponding to $t_1$ and $t_2$. Then, if no structure $f(i_1, i_2): i$ already exists for some $i$, such a structure is created with a novel identifier $i$. In any case, the identifier $i$ (of the new or old structure $f(i_1, i_2): i$) is returned. We also use the following notation: Assume that a term $t$ is represented in a collection of structures. Then, we use $\ida{t}$ to denote the identifier $i$ returned by \textit{toSet} applied to $t$. The value of $\ida{t}$ is not defined, otherwise.
The  operation \textit{substitute} takes as input two identifiers $k$ and $\ell$ and a collection of structures $E$ that uses $k$ and $\ell$. It is not assumed that $E$ is normalized. It removes from $E$ every structure that involves $\ell$ (i.e., a structure of one of the three forms $f(i_1, i_2): \ell$ or $f(\ell, i_2): i $ or $f(i_1, \ell): i$, for some $i_1, i_2, i$) and it adds to $E$ every structure obtained by replacing $k$ by $\ell$ in the previous one, if it is not already in it. 
The operation \textit{normalize} normalizes a collection of structures $E$ by repeatedly applying the operation \textit{substitute} until the collection is normalized. Assuming that it is not, it contains at least two structures $f(i_1, i_2):k$ and $f(i_1, i_2):\ell$, with the same key; the operation \textit{substitute} is thus applied to such $k$, $\ell$. Then, the operation \textit{normalize} is applied to the modified collection. For a precise semantic characterization of the effect of \textit{substitute} and \textit{normalize}, see \cite{atindehouPhDThesis,DBLP:conf/sycss/CharlierA16}.
The last operation, called \textit{unify}, takes as input two identifiers $i$ and $j$, and a collection of structures $E$. It first applies the operation \textit{substitute} to $i$, $j$ and $E$; afterwards, it applies the operation \textit{normalize} to the resulting collection of structures. Semantically, an equivalence constraint is added beween the terms of $T_i$ and $T_j$, and the logical consequences of this constraint are propagated in the whole collection. More precisely, let $\eqrel$ be the abstract denotation of $E$ before applying the operation. After applying it, the collection denotes the smallest congruence $\eqrel'$ such that $\eqrel\:\: \subseteq\:\: \eqrel'$ and $t_i \eqrel' t_j$ (for some (and, in fact, any) $t_i \in T_i$ and $t_j \in T_j$). See \cite{atindehouPhDThesis} for a detailed proof.

The usefulness of collections of structures is greatly due to the fact that they are efficiently implementable (see \cite{atindehouPhDThesis,DBLP:conf/sycss/CharlierA16}). The implementation notably maintains a list \texttt{idList} of all set identifiers and three sets of lists, namely $\texttt{sameId}(i)$, $\texttt{sameId}_1(i_1)$, and $\texttt{sameId}_2(i_2)$, allowing us to go through all structures  $f(i_1, i_2):i$ for $i$, $i_1$, or $i_2$ fixed, respectively. Elements can be added or removed to/from those lists in constant time. The implementation also includes an incremental algorithm that maintains the size of minimal terms in each $E_i$ (denoted by $\textit{size}(i)$).

Previous applications of collections of structures have been described in \cite{atindehouPhDThesis,DBLP:conf/lpar/CharlierA15,DBLP:conf/sycss/CharlierA16}. For instance, it is shown in \cite{DBLP:conf/sycss/CharlierA16} that they can be used to elegantly and efficiently solve the word problem \cite{wechler1992universal} in a theory defined by a finite set of ground equations (see \cite{Downey:1980:VCS:322217.322228,DBLP:conf/stoc/Kozen77,DBLP:journals/jacm/NelsonO80,DBLP:journals/cacm/Shostak78}). This amounts to \emph{solving} all equations $s=t$ in turn, by applying operation \textit{toSet} to $s$ and $t$, and applying operation \textit{unify} to the returned identifiers $i_s$ and $i_t$. Then, to check whether two terms are equal, we only have to apply the operation \textit{toSet} to them and to check whether the returned identifiers are equal. The solution to another, more difficult, problem has been first described in \cite{DBLP:conf/lpar/CharlierA15}, and proven correct in \cite{atindehouPhDThesis}: Collections of structures can be used to compute the (representation of a) congruence defined by a finite set of non ground equations (also called \emph{axioms}) and a finite set of constants, conditionally to the fact that the number of equivalence classes of the congruence is finite (and also, in practice, not too big). The algorithm that solves this problem uses  \emph{valuations}, which are functions from a finite set of variables to the set of all identifiers. Note that we use the letters $x$, $y$, $z$ for variables. Other letters such as $a$, $b$, \ldots\  are used for constants. We generalize the operation \textit{toSet} to non ground terms with an additional argument, namely a valuation whose domain contains the variables of the term. The operation works recursively as before except for variables, in which case the value of the valuation for the variable is  returned. Similarly, we extend the solving of equations as follows:
We apply the operation \textit{toSet} to the left and right sides of the axiom and to some valuation, which returns two identifiers $i_\ell$ and $i_r$. Then, we apply the operation \textit{unify} to $i_\ell$ and $i_r$. In the following, we say that we \emph{apply the valuation to the axiom}. The effect is strictly equivalent to choosing two terms $t_{i_\ell}$ and $t_{i_r}$, represented by $i_\ell$ and $i_r$, and solving the ground equation $t_{i_\ell}\:=\:t_{i_r}$.
The algorithm that computes the congruence starts from a collection of structures representing only the constants. Then it \emph{fairly} generates all valuations that use the variables in the equations and the identifiers of the collection of structures, and applies them to relevant axioms. New identifiers introduced by axiom applications are used in turn to generate new valuations. The algorithm stops when no new valuation can be generated. We call this algorithm the \emph{bottom-up} algorithm. The work presented in the rest of this paper uses the same concepts and operations.

\section{Related Work}
\label{sect:related}

Collections of structures are strongly related to algorithms to compute the \emph{congruence closure} of a relation over terms (see, e.g., \cite{DBLP:conf/tacas/BarbosaFR17,DBLP:books/daglib/0019162,Downey:1980:VCS:322217.322228,kapur1997shostak,%
DBLP:journals/jacm/NelsonO80,NO:RTA-2005,DBLP:journals/cacm/Shostak78}). A main difference is that these previous methods actually work on terms, not on structures as we do, and they use a Union-Find data structures \cite{DBLP:journals/cacm/GallerF64,Knuth:1997:ACP:260999} to record equivalence between terms. Terms often are implemented by DAGs \cite{DBLP:books/daglib/0019162}.
Collections of structures, in some sense, are a generalization of DAGs: When a collection of structures represents a single term, it is represented as a DAG. But, in general, an identifier $i$ represents several, possibly infinitely many, terms. In fact, most congruence closure methods use so-called \emph{signatures} to help detecting equivalence of terms. Our structures are in fact equivalent to signatures: We work with the signatures only and get rid of individual terms. See \cite{atindehouPhDThesis,DBLP:conf/sycss/CharlierA16} for a much more complete comparison.

Computing a congruence closure is often used to build a set of (ground) \emph{rewrite rules} that can be used to simplify terms in theorem provers \cite{kapur1997shostak,DBLP:journals/toplas/NelsonO79,NO:RTA-2005,DBLP:journals/jsc/Snyder93}. Collections of structures can also be viewed as a confluent set of rewrite rules (when the collection is normalized).
Our identifiers are similar to the new constants used in \cite{kapur1997shostak,NO:RTA-2005} and our operation \textit{toSet} simply applies the rewrite rules to a term. However, our ``set of rewrite rules'' is not exactly the same as in \cite{kapur1997shostak,NO:RTA-2005} because we do not use the trick of considering identifiers as new constants.

Collections of structures can also be viewed as a simple form of \emph{tree automata} \cite{tata2007}. The main difference lies in the operations defined on them. The implementation of collections of structures is also specific.

Our approach to expression simplification can be viewed as an instance of \emph{term indexing}~\cite{Schulz:LPAR-2013,Sekar:2001:TI:778522.778535}: in our case, the index is the collection of structures $E$ and the set of indexed terms is its denotation $T$. The relation $R(l, t)$ between an indexed term $l$  and  a query $t$ is that $l$ is a simplification of $t$. This relation is very different in nature from relations such as generalization or instantiation that are usually considered in classical term indexing, but some methods from this area also apply to equational theories such as $AC$ theories (see \cite{Sekar:2001:TI:778522.778535}). 

In the rest of this paper, we illustrate our method by applying it to the simplification of boolean expressions. A lot of work has been done in that area (see e.g., \cite{DBLP:journals/integration/Coudert94,karnaugh1953map,BLTJ:BLTJ3835,quine1955way}) but it mainly concentrates on two-level formulas for logic circuit design. 
BDDs \cite{DBLP:journals/csur/Bryant92} can also be used to represent large boolean formulas compactly, but such a representation is not simple in our sense, i.e., clear and readable.  Our goal is not to compete with such methods. We use the boolean calculus to explain our method mainly because it allows  us to do it very clearly and nicely. Simplifying boolean formulas also is related to SAT solving \cite{DBLP:conf/sat/Knuth12}. Again, our goal is different since SAT solving is a decision problem. However, it could be possibly interesting to simplify input formulas of SAT solvers before they are transformed to equisatisfiable formulas \cite{DBLP:books/daglib/0019162}.
Moreover, application of our method to the simplification of boolean expressions naturally induces effects similar to SAT solving techniques, such as unit clause elimination.


\section{A Method to Simplify Expressions}
\label{sect:method}

\subsection{Principle of the method}
\label{subsec:principle}
The starting idea for our method is as follows: Given an expression to simplify, we could apply all axioms to all its sub-expressions (including itself) in order to build a (possibly large) set of expressions that are equivalent to the original expression. Then, we could either choose a minimal expression within that set, or iterate the process of applying axioms to the newly created sub-expressions. Moreover, having chosen a minimal expression, we can restart the whole process with this expression. And we could continue this way as long as we want, and stop when the current minimal expression would be found satisfactory.

However, in general, the above suggested method is not actually applicable for time and space reasons: Generating all equivalent sub-expressions and keeping them in memory is too inefficient. This is precisely the reason why the collection of structures notion was proposed in \cite{atindehouPhDThesis,DBLP:conf/lpar/CharlierA15,DBLP:conf/sycss/CharlierA16}:
Collections of structures allow us to manipulate sets of equivalent terms globally without enumerating them. For boolean expressions, using suitable axioms, the bottom-up algorithm explained at the end of Section \ref{sect:structures} computes a representation of all possible expressions (with $n$ constants) and, in a sense, it solves all instances of the simplification problem, since any expression can be instantly simplified by applying the \textit{toSet} operation to it (in the context of the collection of structures representing the congruence). Unfortunately, the number of structures $N$ needed for the collection is proportional to $2^{2^n}$ where $n$ is the number of constants. Thus, the method works only for very small values of $n$.
 For simplifying expressions using more constants, our idea is to adapt the bottom-up algorithm by focusing on the expression to be simplified. We limit the generation of valuations and the application of axioms to cases where they are relevant for the task of simplifying this particular expression.

Remember that, in the case of the bottom-up algorithm, we apply a valuation to an axiom as follows:  We apply the operation \textit{toSet} to the left and right sides of the axiom and to some valuation, which returns two identifiers $i_\ell$ and $i_r$. Then, we apply the operation \textit{unify} to $i_\ell$ and $i_r$. To focus on a particular expression, we apply the operation \textit{unify} only if at least one of the two identifiers is in the collection of structures beforehand. This ensures that the collection of structures only represents expressions equivalent to the expression to be simplified (and, of course, sub-expressions of these expressions). 
%
%
For valuations, we would ideally only generate useful valuations that add terms or equality constraints to the collection of structures, when applied to some axioms.
We have first attempted to compute such valuations by implementing a form of pattern matching between the terms in the axioms and the structures of the collection. It happens that this method is impracticable (i.e, both complicated and inefficient) because identifiers and structures of a collection potentially represent many terms, not a single one, and because there can be a lot of redundancy induced by the matching operation: many returned valuations can be the same. Fortunately, we have found a way to generate interesting valuations in a much simpler way: We consider all identifiers $i$ of the collection, in turn, in order of appearance in the collection. We execute all axioms involving a single variable $x$ with respect to the valuation $\{ x \mapsto i\}$. We run through all structures $f(j_1, j_2):i$ (for this particular $i$) and we execute all axioms involving exactly two variables $x$, $y$, with respect to the valuation $\{ x \mapsto j_1,\ y \mapsto j_2\}$. For the axioms involving three variables $x$, $y$, $z$, we then consider all structures of the form $f'(k_1, k_2): j$ where $j\in\{j_1, j_2\}$ and execute the axioms with respect to the valuation $\{ x \mapsto i_1,\ y \mapsto i_2,\ z \mapsto i_3 \}$ where $\langle i_1, i_2, i_3 \rangle = \langle k_1,  k_2,  j_2 \rangle$ (and $j = j_1$) or $\langle i_1,  i_2,  i_3 \rangle = \langle j_1, k_1, k_2 \rangle$ (and $j = j_2$). Note that we assume that axioms use at most three variables. More elaborated valuations should be used for more complex axioms (see Section \ref{subsec:variants}). 

The reader may wonder in what sense generating valuations and applying them to axioms, as explained above, can actually simplify an expression. In fact, it just effectuates, on collections of structures, the kind of treatment we suggested, on sets of terms, at the beginning of this section. We can now make it clearer with an example.


\begin{figure}[h]
\caption{Axioms for Simplifying Boolean Expressions}
$$\begin{array}{|crclc|crclc|crclc|crclc|}\hline
		&&&&&&&&&&&&&&&&&&&\\[-0.5em]
&x\ 1  & = & x   &&& 1 + x  & = & 1   &&&
 xy    & = & yx  &&& (x + y) + z  & = & x + (y + z)  &\\[0.2em]

&1\ x  & = & x   &&& x + 1  & = & 1   &&&
 x + y  & = & y + x  &&& (x + y)z  & = & xz + yz  &\\[0.2em]

&x\ 0  & = & 0   &&& \overline{x}\ x  & = & 0   &&&
 (xy)z    & = & x(yz) &&& x(y + z)  & = & xy + xz  &\\[0.2em]

\cline{16-20}
&&&&&&&&&&&&&&&&&&&\\[-0.9em]
&0\ x  & = & 0   &&& x + \overline{x}  & = & 1 &&&
 \overline{xy}  & = & \overline{x} + \overline{y} &&&
x\, y \,z\, \overline{y} &=& 0 &\\[0.2em]

&x + 0  & = & x   &&& \overline{x} + x  & = & 1 &&&
 \overline{x + y}  & = & \overline{x}\  \overline{y} &&&
x + y + z + \overline{y} &=& 1  &\\[0.2em]

&0 + x  & = & x   &&& \overline{\overline{x}}  & = & x   &&&
 (x + y)z  & = & xz + yz &&&
(x + y + z) y &=& y &\\[0.2em]
 
&&& &&&               x\ \overline{x}  & = & 0   &&&
 x(y + z)  & = & xy + xz &&&
x\,y\,z + y & = & y &\\[0.2em]

		\hline
\end{array}$$
\label{fig:axioms}
\end{figure}

%
\begin{example}
\label{lab:ex2}
Let us apply our method to simplify the boolean expression \texttt{a + b + !b + a} using the axioms of Figure \ref{fig:axioms}. Note that we write \texttt{!b} instead of $\overline{b}$, as in Figure \ref{fig:axioms}. In order to very concretely show what happens, we use the actual numerical values of the identifiers. So, after creating the representation of \texttt{a + b + !b + a} in the collection of structures (i.e., after applying the operation \textit{toSet} to it), we have the following values for the identifiers:
$$
\begin{array}{rclcrclcrclcrclcrcl}
\ida{0} &=& 1,&\hspace*{0.4em}& \ida{1} &=& 2,&\hspace*{0.4em}& \ida{a} &=& 3,&\hspace*{0.4em}& \ida{b} &=& 4,\\ \ida{a + b} &=& 5,&\hspace*{0.4em}& \ida{!b} &=& 6,&\hspace*{0.4em}& \ida{a + b + !b} &=& 7,&\hspace*{0.4em}& \ida{a + b + !b + a} &=& 8.
\end{array}
$$
In other words, the following structures exist:
$$
\begin{array}{ccccccccccccccccc}
\texttt{0}:1,  &\hspace*{0.7em}&  \texttt{1}:2, &\hspace*{0.7em}&   \texttt{a}:3,  &\hspace*{0.7em}&  \texttt{b}:4,   &\hspace*{0.7em}& 
+(3, 4):5,  &\hspace*{0.7em}&  !(4):6,  &\hspace*{0.7em}& +(5, 6):7,  &\hspace*{0.7em}& +(7, 3):8.          
\end{array}
$$
Now, we apply the method: we go through the list of identifiers, compute the valuations related to each identifier, and apply the valuations to the axioms. Consider the moment when identifier $7$ is taken into account. The valuation $\val = \{x \mapsto 3, y \mapsto 4, z \mapsto 6\}$ is generated because the structures $+(3, 4):5$ and $+(5, 6):7$ exist. 
Thus, the operation $\toSet$ is applied to the term  $(x + y) + z$, which returns the identifier $7$, thanks to the very same structures. So the operation $\toSet$ is also applied to the term $x + (y + z)$. We first get: $\toSet(y + z,\ \val) = 2$, because the axiom $x + \overline{x} = 1$ has been applied to the valuation $\{x \mapsto 4\}$, previously; so, the structure $+(4, 6):2$ exists.
Then, we get $\toSet(x + (y + z),\ \val) = 2$, because the structure $+(3, 2):2$ exists, due to a previous application of the valuation $\{x \mapsto 3 \}$ to the axiom $x + 1 = 1$. To conclude the axiom application, identifiers $7$ and $2$ are unified: First, $7$ is replaced by $2$ in the collection of structures, changing the structures $+(5, 6):7$ and  $+(7, 3):8$ into $+(5, 6):2$ and  $+(2, 3):8$. But, at this point, the collection of structures also contains the structure $+(2, 3):2$, due to a previous application of the axiom $1 + x = 1$. Therefore, the operation \textit{normalize} unifies the identifiers $8$ and $2$, replacing $8$ by $2$. Since $8$ was the identifier of expression  \texttt{a + b + !b + a}, this identifier is now set to $2$. In other words, the expression \texttt{a + b + !b + a} has been simplified to \texttt{1}. And this has been achieved by a single axiom application (which takes advantage of a lot of work done before, however).

Observe that the axiom application has used the constant $1$ (in fact, its identifier) \emph{three times}, despite that the identifier of $1$ is not used by the valuation \val. Moreover, the result of unifying $7$ and $2$ is immediately propagated in the collection of structures, so that no additional computation remains needed. We say that this axiom application \emph{uses} $1$.
\end{example}
More generally, we say that \emph{an axiom application uses $0$ or $1$} if one or both of their identifiers appears at some point of the computation of this axiom application or in the valuation applied to the axiom. This kind of axiom application plays a major role in the simplification process (see Section \ref{sec:experiments}).\footnote{Another example, where $1$ is used by the valuation, is given in the optional appendix.} 

\begin{figure}[ht!]
\caption{The algorithm in principle}
\vspace{-0.2cm}
\begin{center}
\begin{tabular}{|c|}\hline
\\[-0.8em]
\seq{
   \ Create an empty collection of structures; set \texttt{mainId} to \textit{toSet}(\texttt{expr});\\
   \ set \texttt{previousSize} to \textit{size}(\texttt{mainId});
   set \texttt{count} to $0$;\\[0.2em]

   \while{9}{\texttt{count} $\neq$ \texttt{maxCount} \textbf{and} \textit{size}(\texttt{mainId}) $>$ \texttt{expectedSize}\\}
   {  
   
   \ set \texttt{reCount} to $0$;\\
   \while{9.3}{\texttt{reCount} $\neq$ \texttt{maxReCount} \textbf{and} the collection of structures is not full\\}
    {  \seq{reset \texttt{idList}; set \texttt{timeLimit} to current time;
       set \texttt{subCount} to $0$;}\\[0.2em]
      \while{10}{\texttt{idList} is not fully traveled
            \textbf{and} the collection of structures is not full\\
            \textbf{and} \texttt{subCount} $\neq$  \texttt{maxSubCount}
            \textbf{and} \textit{size}(\texttt{mainId}) $>$ \texttt{expectedSize}\\

            }
       { \ let $i$ be the next identifier in \texttt{idList};\\
         \ compute the valuations related to $i$;\\
         \while{5}{the set of valuations is not empty\\}
         {
           pick a valuation \textit{val} in the set;\\
           apply \textit{val} to all axioms that have the same arity as \textit{val};
         }\\[2em]
       
         \myif{4}{ \textit{time}($i$) $>$ \texttt{timeLimit}}
         {
           add 1 to \texttt{subCount};\\
           set \texttt{timeLimit} to current time;
         }\\
       }\\ \ \\[-0.5em]
       \,add 1 to \texttt{reCount};
       
       }

         \\\ \\[-0.5em] \ call garbage collector;\\[0.5em]
         
         \myif{4.8}{\textit{size}(\texttt{mainId}) $=$ \texttt{previousSize}}
         { add $1$ to \texttt{count};} \\[2em]
         
         \myif{4.8}{\textit{size}(\texttt{mainId}) $<$ \texttt{previousSize}}
         { reset \texttt{count} to 0;\\
           set \texttt{previousSize} to \textit{size}(\texttt{mainId});
         }         
   }
}\\[-0em]
\\\hline
\end{tabular}
\end{center}
\label{fig:algo}
\end{figure}
We are now almost in position to present a first simple algorithm that implements our method. 
The only remaining issues are to decide what to do when the list of identifiers initially present in the collection of structures is completely visited, and to decide what to do when the memory assigned for the collection of structures is full. For the first issue, we may choose to  continue by visiting the identifiers newly created up to now, or by restarting an iteration from the beginning. The first strategy, in some sense, amounts to performing a depth-first search in the set of newly created terms, while the second strategy resembles a breadth-first search. No strategy is a priori ideal. Therefore, the algorithm uses two counters \texttt{reCount} and \texttt{subCount} to allow a compromise between the two: the first strategy is used until \texttt{subCount} reaches a maximum value \texttt{maxSubCount}; then, the computation is restarted at the beginning of the identifier list (\texttt{idList}) at most \texttt{maxReCount} times. As for the second issue, the collection of structures may become full at any moment when processing the list of identifiers. At that point, a form of garbage collection is applied. The main contract that garbage collection must respect is to keep enough structures for representing at least one of the minimal expressions represented at call time. 
The counter \texttt{reCount} is also needed to ensure termination when a fixpoint (stable) collection of structures is obtained before a garbage collection call is needed.
A simple version of the algorithm is depicted in Figure \ref{fig:algo}. 
The expression $\textit{time}(i)$ returns the value of the clock maintained by the collection of structures when the first structure $f(i_1, i_2):i$ was created. When two identifiers are unified we always replace the younger by the older. The variable \texttt{idList} maintains the current list of identifiers in the collection of structures, sorted in chronological order. (Renamed identifiers are automatically removed from the list by the \textit{unify} operation. Similarly, garbage collection cleans up the list, keeping only identifiers remaining in the collection of structures.) Finally, termination is ensured by counting how many times an iteration has been performed without making the size of the smallest expression decrease. One can also specify an \texttt{expectedSize} for the minimal expression to possibly avoid useless iterations. 
The algorithm starts by creating an empty collection of structures,
applying the operation \toSet\ to the expression \texttt{expr} to be simplified, and setting the variable \texttt{mainId} to the value of its identifier.

\subsection{Difficult issues, solutions and workarounds}
\label{subsec:workaround}

Experiments with the algorithm of Figure \ref{fig:algo} have revealed several weaknesses: A major problem arises when an \emph{an iteration}, i.e, any execution of the body of the main loop of the algorithm, up to the garbage collector call 
is unable to consider each and all of the identifiers existing at the beginning of the iteration. This leads to what we call \emph{an early-fixpoint}: the algorithm stops after simplifying some sub-expressions well but ignoring completely parts of the whole collection of structures. The major cause of the problem is a kind of unfairness: new structures are generated by applications of the axioms, and they can possibly be taken into account immediately for generating new valuations. We solve this issue by only considering structures that have been created before the current \emph{sub-iteration} 
has started, i.e, before the last time that the variable \texttt{timeLimit} has been changed. It is also useful to limit the size of the identifiers in the valuations, as well as the size of the structures created by axiom applications. The general rule is that a structure $f(i_1, i_2):i$ is acceptable for generating a valuation only if $\textit{size}(i_j) \leq \textit{size}(i)$ ($j=1,2$). Similarly, such a structure can be created only if  $\textit{size}(i_j) \leq \textit{size}(i')$ ($j=1,2$) where $i'$ is the existing identifier which is about to be unified with $i$. But there are still situations where the proposed improvements are not powerful enough. Therefore, we have introduced additional workarounds that allow the algorithm to consider more identifiers, and to stops within predictable time. The ultimate and most drastic of these consists of adding a time-out to each iteration.\\[-2em]


\subsection{Variants of the method}
\label{subsec:variants}

There are still some decisions underlying our algorithm that are not completely made explicit. Moreover, some variants are interesting to investigate. 
\begin{itemize}[leftmargin=0in]
\item 
The kind of valuations we have proposed may be not powerful enough to ensure that all interesting axiom applications are performed.  So, we have tried three other \emph{kinds of valuations} that we identify by a type number between 0 and 3. The ones we have used up to now are given type 0. Valuations of type 3 include all combinations of existing identifiers. They are generated as follows: each time an identifier $i$ is taken into account by the algorithm (see Figure \ref{fig:algo}), all valuations $\{x \mapsto i\}$, $\{x \mapsto i, y \mapsto i'\}$, and 
$\{x \mapsto i, y \mapsto i', z \mapsto i''\}$, with $time(i) \ge time(i') \ge time(i'')$ are considered for axiom application. This type of valuations is more complete, even exhaustive when multiple axiom application is used (see below), but they more often provoke the early-fixpoint phenomenon. Valuations of types 1 and 2 are somehow intermediate between 0 and 3. They allow more combinations of identifiers but they must stay ``close'' to $i$ in the collection of structures.
\item 
When a valuation is applied to an axiom, we allow either a \emph{strict (unique) application} of this particular valuation or \emph{multiple application} by all valuations obtained by exchanging identifiers between variables in the given valuation. In case of multiple application we normalize the valuations to avoid redundancy. 
\item As we have seen, we normally use conditional axiom application. 
Alternatively, we may choose to apply the axioms freely, without  any precondition. Structures $f(i_1, i_2):i$ may be created that are not related to the expression to simplify, at this time. But further axiom applications may later unify $i$ with another identifier that is actually used to represent the current minimal expression. Thus, this gives us a chance to create more interesting structures, in the long run. We call this the \emph{bottom-up application} of axioms. This method requires us to apply another kind of garbage collection when the memory is full: We only remove structures that are not reachable from the \texttt{mainId} identifier.
\item 
Many logically equivalent \emph{axiom sets} can be chosen for any equational theory. When used by a version of our simplification algorithm, their practical value can be very different. Adding more axioms can make the simplification process faster sometimes, but executing more axioms takes more time and it can create more structures, leading to an earlier garbage collector call. 
\item 
We have seen in Example 
  \ref{lab:ex2} 
that axioms involving a single variable are especially useful because they create structures containing the identifiers \ida{0}\ or \ida{1} that are later exploited by other axioms.
This suggests that, at each iteration, or sub-iteration, we first apply all axioms involving a single variable to all identifiers to which they have not been already applied. We call this the \emph{early-application} of these axioms. Let us slightly change Example \ref{lab:ex2} by considering the expression \texttt{a + b + !b + c}
. Normal application of the axioms simplifies it to \texttt{1 + c}, not to \texttt{1}, because the identifier \ida{c} has not yet been processed by the algorithm. If we use early-application, the structure $+(\ida{1}, \ida{c}):\ida{1}$ already exists, so that the expression is simplified to \texttt{1}. In fact, this is a general rule: If early-application is used, then a current minimal expression never contains one of the two symbols \texttt{0} or \texttt{1}, except if it is equal to one of them.
\end{itemize}
\ \\[-4em]

\section{Experimental Evaluation}
\label{sec:experiments}

\vspace{-1em}We now give an experimental evaluation of our simplification method, applied to boolean expressions. First, we provide 
statistics about its application to a set of randomly generated expressions. Second, we analyze the results of 13 variants of the algorithm for 7 particular expressions. Third, we delve deeper into the execution of 
the algorithm on specific expressions
.

\begin{table}[t!]
\caption{Statistics on simplifying expressions}
\vspace{-0.5cm}
$$\begin{array}{|r||r|r||r|r||r|r||r|r||r|r||}\hline
\textrm{Size}  &  \multicolumn{2}{c||}{\textrm{3 letters}} &  \multicolumn{2}{c||}{\textrm{5 letters}} &  \multicolumn{2}{c||}{\textrm{7 letters}} &  \multicolumn{2}{c||}{\textrm{9 letters}} &  \multicolumn{2}{c||}{\textrm{16 letters}} \\\hline\hline
 1  & 58 & 0.10 & 43 & 0.51 & 29 & 1.06 & 28 & 1.25 & 19 & 4.53\\
5 & 75 & 0.11 & 47 & 0.56 & 30 & 1.08 & 31 & 1.26 & 19 & 4.53\\
6 & 92 & 0.11 & 51 & 0.70 & 31 & 1.09 & 31 & 1.26 & 19 & 4.53\\
10 & 97 & 0.11 & 77 & 1.72 & 36 & 1.41 & 33 & 1.26 & 20 & 4.4\\
15 & 99 & 0.12 & 87 & 1.98 & 55 & 2.09 & 41 & 1.56 & 21 & 4.34\\
16 & 100 & 0.13 & 88 & 2.05 & 57 & 2.29 & 44 & 1.65 & 21 & 4.34\\\hline
19 &  &  & 98 & 2.75 & 63 & 2.55 & 50 & 2.05 & 23 & 4.80\\
24 &  &  & 100 & 2.82 & 73 & 2.92 & 56 & 2.50 & 26 & 7.47\\\hline
26 &  &  &  &  & 76 & 2.99 & 60 & 3.40 & 30 & 7.78\\
39 &  &  &  &  & 93 & 6.83 & 75 & 7.73 & 36 & 11.37\\
57 &  &  &  &  & 100 & 8.68 & 87 & 9.56 & 47 & 14.45\\\hline
64 &  &  &  &  &  &  & 89 & 10.68 & 50 & 14.89\\
135 &  &  &  &  &  &  & 100 & 12.94 & 71 & 21.93\\\hline
157 &  &  &  &  &  &  &  &  & 75 & 25.49\\
318 &  &  &  &  &  &  &  &  & 100 & 30.03\\\hline
\hline
\textrm{Average} & 3.12 & 0.13 & 6.91 & 2.82 & 16.39 & 8.68 & 27.15 & 12.94 & 94.59 & 30.03\\\hline
\end{array}$$
\label{tab:tailles}
\end{table}

In the first part of this evaluation, we use the default version of our algorithm (see Table \ref{lab:synthesisCasDeTests}). We have randomly generated five sets of 100 boolean expressions, built of the symbols '.', '+', and '!', and of 3, 5, 7, 9, or 16 different letters (i.e., constants). Each expression has a size equal to 800. The size of an expression is the number of symbols needed to write it Polish notation. 
We have applied the algorithm to every such expression, and collected the size of its simplified version as well as the time needed to compute it. Statistics on this experiment are given in Table \ref{tab:tailles}.
In the first column, we give some possible sizes for the simplified expressions. Corresponding to each such size, we provide in each group of two columns the number of simplified expressions that have at most this size and the average time needed to compute them (in seconds). The last line depicts the average size of a simplified expression and the average computation time. Among other things, we see that many expressions are simplified to a single symbol (most often \texttt{0} or \texttt{1}), but their number decreases with the number of letters in the original expression. Also, the simplification time increases with the size of the simplified expression
. 
Note that the reported times are not the actual times spent by the algorithm until termination (except in the case where the minimal size is 1): When the final size is reached, the algorithm still performs twenty iterations -- hoping for further simplifications -- before it stops. 

We now turn to a comparison of different variants 
of our simplification algorithm. For this comparison, we have selected five typical expressions ($\textrm{E} 3$ to $\textrm{E} 16$) of size 800, using  3, 5, 7, 9, and 16 letters,  
as well as two ``very big'' expressions ($\textrm{B} 3$ and $\textrm{B} 9$) of size $100,000$, using 3 and 9 letters.\footnote{For the reviewers: See the optional appendix for more information.}
The variants of our algorithm are described in Table \ref{lab:synthesisCasDeTests}.
\begin{table}[t!]
\caption{Description of 13 variants of the algorithm}
\vspace{-0.5cm}
$$\begin{array}{|l||c|c|c|c|c|c|}\hline
\textrm{Variant} & \textrm{Valuations} &  \textrm{Appl. Mode} & \textrm{maxSubCount}  & \textrm{BU?} &  \textrm{Axioms} & \textrm{Algorithm} \\\hline\hline
\textrm{default} & 0 &  & 6 & &  &  \\
\hline
\textrm{var/}1 & 0 & \textrm{U} & 6 & &  &  \\
\textrm{var/}2 & 0 &  & 1 & &  &  \\
\textrm{var/}3 & 0 &  & 6 & \textrm{BU}&  &  \\
\textrm{var/}4 & 0 &  & 6 & & \textrm{A+} &  \\
\textrm{var/}5 & 0 &  & 6 & &  & \textrm{1F} \\
\textrm{var/}6 & 0 & \textrm{U} & 1 & &  &  \\
\hline
\textrm{var/}7 & 1 &  & 6 & &  &  \\
\textrm{var/}8 & 1 &  & 1 & & \textrm{A+} &  \\
\hline
\textrm{var/}9 & 2 &  & 6 & &  &  \\
\textrm{var/}10 & 2 & \textrm{U} & 1 & &  &  \\
\hline
\textrm{var/}11 & 3 &  & 6 & &  &  \\
\textrm{var/}12 & 3 &  & 1 & &  & \textrm{1F} \\
\hline\end{array}$$
\label{lab:synthesisCasDeTests}
\end{table}

The first variant, called ``default'', is the version used for the first experiment. It proceeds exactly as explained in Section \ref{subsec:principle} and Section \ref{subsec:workaround}: It uses the standard way to generate valuations (0), and multiple application of the axioms to valuations; it limits the number of sub-iterations to 6, uses conditional axiom application, uses the standard set of axioms, and applies axioms to valuations in the normal way. 
The other variants use one or more of the ideas explained in Section \ref{subsec:variants}: The second column indicates the kind of valuations that are used. A `U' in the third column means that application of valuations to axioms  is strict (unique: no permutation is made inside valuations); otherwise, application is multiple. A `1' in the fourth column says that an iteration is limited to a single sub-iteration, but it can be repeated up to $3$ times (we set \texttt{maxReCount} to $3$). `BU' in the fifth column means that bottom-up axiom application of axioms is used. When `A+' is put in the sixth column, the set of axioms is extended with the four non standard axioms at the bottom right of Figure \ref{fig:axioms}. Finally, mentioning `1F' in the seventh column tells us that, in every sub-iteration, axioms using only one variable are all first executed for all identifiers (to which they have not been previously applied) before the axioms using more than one variable are taken into account. 
Every selected variant differs from the default algorithm
by only one feature, except variants $\textrm{var/}8$, $\textrm{var/}10$, and $\textrm{var/}12$, which have been chosen among the best variants using valuations of type 1, 2, or 3, respectively. Variant $\textrm{var/}6$ is an additional choice, which is particularly fast in a single test case ($\textrm{E}7$). 
Results are given in Table \ref{lab:synthesisSuperTests}. (Best results are in bold.) We can make the following comments. 
\begin{table}[t!]
\caption{Comparison of 13 variants of the algorithm}
\vspace{-0.5cm}
$$\begin{array}{|l||r|r||r|r||r|r||r|r||r|r||r|r||r|r||}\hline
\textrm{Variant} &  \multicolumn{2}{c||}{\textrm{E} 3} &  \multicolumn{2}{c||}{\textrm{E} 5} &  \multicolumn{2}{c||}{\textrm{E} 7} &  \multicolumn{2}{c||}{\textrm{E} 9} &  \multicolumn{2}{c||}{\textrm{E} 16} &  \multicolumn{2}{c||}{\textrm{B} 3} &  \multicolumn{2}{c||}{\textrm{B} 9} \\\hline\hline
\textrm{default}
 & 1 & \bf 0.10 & 11 &  0.30 & 22 & 10.12 & 24 & \bf 16.53 & 52 & \bf 9.73 & 6 & \bf  0.66 & \bf 49 & 73.44\\
\hline
\textrm{var/}1
 & 1 & 0.15 & 11 & \bf 0.27 & 22 & 7.27 & 24 & 80.99 & 52 & 51.52 & 6 & 0.70 & 50 & 133.29\\
\textrm{var/}2
 & 1 & 0.20 & 11 & 0.58 & 22 & 3.55 & 25 & 28.59 & 52 & 51.49 & 6 & 1.38 & 50 & 477.97\\
\textrm{var/}3
 & 1 & 0.17 & 11 & 1.17 & 22 & 37.11 & 25 & 181.40 & 52 & 178.36 & 6 & 0.86 & 50 & 422.17\\
\textrm{var/}4
 & 1 & 0.10 & 11 & 0.35 & 22 & 7.24 & 24 & 25.45 & 52 & 11.02 & 6 & 0.77 & 50 & \bf  34.54\\
\textrm{var/}5
 & 1 & 0.14 & 11 & 0.43 & 22 & 14.26 & 24 & 120.27 & 52 & 38.55 & 6 & 1.56 & 50 & 97.06\\
\textrm{var/}6
 & 1 & 0.21 & 11 & 0.73 & 22 & \bf 1.62 & 24 & 34.19 & 52 & 17.29 & 6 & 1.15 & 50 & 391.38\\
\hline
\textrm{var/}7
 & 1 & 0.38 & 11 & 2.52 & 22 & 22.37 & 25 & 163.24 & 52 & 313.62 & 6 & 8.79 & 50 & 654.87\\
\textrm{var/}8
 & 1 & 0.26 & 11 & 2.21 & 22 & 50.41 & 24 & 132.68 & 52 & 156.72 & 6 & 11.63 & 50 & 263.79\\
\hline
\textrm{var/}9
 & 1 & 0.19 & 11 & 2.79 & 22 & 80.90 & 25 & 528.26 & 75 & 530.43 & 6 & 20.12 & 50 & 1334.14\\
\textrm{var/}10
 & 1 & 0.15 & 11 & 0.78 & 22 & 23.65 & 24 & 210.03 & 52 & 528.01 & 6 & 5.04 & 50 & 579.09\\
\hline
\textrm{var/}11
 & 1 & 0.80 & 11 & 7.59 & 22 & 145.68 & 25 & 513.24 & 52 & 1789.17 & 6 & 6.39 & 73,515 & 1740.00\\
\textrm{var/}12
 & 1 & 0.39 & 11 & 8.70 & 22 & 36.41 & 24 & 120.21 & 52 & 150.12 & 6 & 2.81 & 57,951 & 840.00\\
\hline\end{array}$$
\label{lab:synthesisSuperTests}
\end{table}

\begin{itemize}[leftmargin=0in]
\item Except for three test cases, all variants give precise results (size of minimal expressions) but possibly very different execution times, which shows that it is difficult to find a unique best parametrization for the algorithm. The default algorithm always gives most precise results and it is also fastest in four cases. However, it is far outperformed by $\textrm{var/}6$ on $\textrm{E}7$. It is also largely outperformed by $\textrm{var/}4$ on $\textrm{B}9$ but it gives a slightly more precise result on that test case.
\item When the number of letters 
is small, it is sometimes better to limit the number of sub-iterations, or to use unique instead of multiple axiom application.
Moreover, in that case, all variants give the same minimal size, suggesting that minimizing expressions remains easy. For 9 letters and beyond, it may be very difficult to get an expression of minimal size from an ``almost minimal'' one. For instance, the expression
\texttt{b + (g + a)d + i + !(hfe(d + ag!c))} has a size equal to $25$, and is returned by variant 
$\textrm{var/}2$ for $\textrm{E}9$. It is quite difficult to transform it into 
an expression of size $24$ using the axioms of Figure \ref{fig:axioms}. (The reader should try it.)

\item The variant $\textrm{var/}3$ suggests that applying axioms in the bottom-up way (i.e., freely) is not useful in general since it increases the execution time by up to an order of magnitude.

\item The variant $\textrm{var/}4$ shows that using the extended set of axioms  (A+) does not always decrease the execution time. It does so only in two cases, of which the most interesting is $\textrm{B}9$. 
Using the standard axioms is reasonable, in general.

\item The time efficiency of variants $\textrm{var/}7$ to $\textrm{var/}10$ is rather disappointing. The results are precise however, suggesting that valuations of type $1$ and $2$ could be useful for some applications using other equational axioms.

\item The results for $\textrm{var/}11$ and $\textrm{var/}12$ show that valuations of type 3 (generated in a pure bottom-up way) are 
unable to simplify the very large expression $\textrm{B} 9$: many identifiers of the initial expression are never taken into account and an early-fixpoint collection of structures is obtained. 
The algorithm terminates in reasonable time because of the 60-second time-out applied at each iteration. Otherwise, termination would take a time that we can not even estimate. (The first iteration takes more than one day.)
Nevertheless, the method works reasonably well for smaller expressions. Since this method attempts to generate all possible valuations, the results for $\textrm{var/}12$ suggest that it could be used in contexts where valuations of type 0 are not effective. Note that 
it is always better, for efficiency, to use the algorithm 1F with valuations of type $3$. This is not the case for valuations of type 0, as shown by the results for $\textrm{var/5}$.

\item Finally, we observe that the size of the expression is not the major impediment for the simplification task: the ratios of the execution times for $\textrm{B}3$ and $\textrm{B}9$ by those for $\textrm{E}3$ and $\textrm{E}9$ are much smaller than the ratio of their sizes (125). Actually, for valuations of type 0, 1, and 2, large expressions provide many more valuations than small ones, which makes more simplifications possible. 
\end{itemize}

\begin{table}[t!]
\caption{Execution of the \textrm{default} algorithm  on expression \textrm{E}9}
\vspace{-0.5cm}
$$\begin{array}{|r|r||r|r|r||r|r|r|r||r||r|r|r|r|}\hline
\textrm{time}  & \textrm{size} & \textrm{nval} &  \textrm{napl} & \textrm{a} / \textrm{v} & \textrm{ds01}  & \textrm{nd01} & \textrm{ods} & \textrm{nods} & \textrm{nid}  & M_i & N_i & M_f & N_f \\\hline\hline
1.83 & 143 & 105K & 337K & 3.1 & 530 & 82 & 127 & 66 & 11K & 373 & 372 & 64K & 353K\\
0.26 & 130 & 28K & 111K & 3.9 & 8 & 3 & 5 & 3 & 2.8K & 205 & 540 & 26K & 120K\\
0.27 & 126 & 29K & 117K & 4.1 & 2 & 1 & 2 & 1 & 2.6K & 162 & 491 & 28K & 128K\\
0.65 & 114 & 61K & 259K & 4.2 & 11 & 5 & 1 & 1 & 5.0K & 185 & 628 & 66K & 289K\\
0.19 & 108 & 19K & 82K & 4.3 & 0 & 0 & 6 & 3 & 2.0K & 117 & 326 & 17K & 86K\\
0.26 & 102 & 25K & 104K & 4.2 & 4 & 1 & 2 & 1 & 2.3K & 130 & 374 & 24K & 111K\\
0.28 & 102 & 27K & 117K & 4.3 & 0 & 0 & 0 & 0 & 2.5K & 130 & 393 & 29K & 129K\\
0.58 & 94 & 51K & 211K & 4.1 & 8 & 1 & 0 & 0 & 3.9K & 732 & 2.6K & 57K & 228K\\
0.63 & 61 & 53K & 214K & 4.0 & 29 & 8 & 4 & 2 & 4.4K & 860 & 3.3K & 51K & 218K\\
0.29 & 53 & 26K & 100K & 3.8 & 8 & 2 & 0 & 0 & 2.3K & 477 & 2.0K & 21K & 101K\\
1.24 & 30 & 79K & 365K & 4.5 & 23 & 4 & 0 & 0 & 4.1K & 155 & 477 & 53K & 314K\\
\hline
1.08 & 26 & 82K & 330K & 4.0 & 0 & 0 & 4 & 2 & 3.1K & 108 & 305 & 75K & 358K\\
1.01 & 26 & 79K & 342K & 4.3 & 0 & 0 & 0 & 0 & 10K & 84 & 225 & 23K & 278K\\
3.81 & 26 & 172K & 418K & 2.4 & 0 & 0 & 0 & 0 & 7.4K & 25 & 24 & 31K & 288K\\
2.03 & 26 & 178K & 489K & 2.7 & 0 & 0 & 0 & 0 & 2.6K & 25 & 24 & 75K & 431K\\
0.54 & 26 & 47K & 206K & 4.4 & 0 & 0 & 0 & 0 & 6.6K & 14K & 28K & 27K & 210K\\
1.5 & 25 & 127K & 323K & 2.5 & 0 & 0 & 1 & 1 & 2.6K & 58 & 178 & 75K & 341K\\
0.0 & 24 & 383 & 2.1K & 5.4 & 0 & 0 & 1 & 1 & 96 & 52 & 159 & 559 & 2.6K\\
\hline
\end{array}$$
\label{tab:laba}
\end{table}


We now delve deeper into the execution of the algorithms. 
We give statistics on the execution of the default algorithm for expression $\textrm{E}9$ 
in Table \ref{tab:laba}. Each line of the table gives information about an iteration of the algorithm. For instance, we see in the first column of the first line, that the execution time of the first iteration is $1.83$ seconds. In the second column, we see that the size of a smallest expression at the end of the first iteration is 143. The column $\textrm{nval}$ gives the total number of valuations generated during an iteration, while the column $\textrm{napl}$ gives the number of useful axiom applications. We say that an axiom application is \emph{useful} when it modifies the collection of structures. The column $\textrm{a} / \textrm{v}$ gives the ratio of $\textrm{napl}$ by $\textrm{nval}$. The next four columns help us  understand how the simplification takes place. For instance, the values $530$ and $82$ in columns $\textrm{ds01}$ and $\textrm{nd01}$ indicate that the size of the minimal expression has decreased $82$ times during the first iteration due to an axiom application that uses $0$ or $1$  (remember Example \ref{lab:ex2}), and that the total reduction in size achieved by those applications is equal to $530$. Columns  $\textrm{ods}$ and $\textrm{nods}$ indicate that $66$ other axiom applications have reduced the minimal size by $127$ symbols. Column \textrm{nid} provides the number of times that an identifier $i$ is selected by the algorithm (at the ninth line of Figure \ref{fig:algo}). Columns $M_i$ and $N_i$ respectively give the number of identifiers and the number of structures in the collection at the beginning of an iteration. Columns $M_f$ and $N_f$ are the corresponding numbers at the end of the iteration, just before the garbage collector is called.

We observe that the first iteration provides most of the simplification. Nevertheless, a lot of work is still to be done since an expression of $143$ symbols certainly is much less simple than an expression of $24$ symbols. We also see how important the axiom applications that use $0$ or $1$ are. At the first iteration, they provide $81 \%$ of the size decrease.
Moreover, the average reduction in size for these axiom applications is $6.5$, while
it is only equal to $1.9$ for other (size reducing) axiom applications.
During the next iterations, the decrease slows down and a plateau appears at size $102$, but during the next four iterations the decrease becomes substantial anew, mostly due again to axiom applications using $0$ or $1$. Afterwards, we have drawn a horizontal line in the table to stress the fact that these axiom applications are no longer useful, i.e., do not happen anymore. Simplification becomes now a purely combinatorial issue. A plateau of length $4$ is traversed before the last two iterations finish the simplification. Some iterations in the plateau take significantly more time than the ``productive''  iterations: the algorithm is working very hard for apparently nothing. Nevertheless, the way it ``shakes'' the collection of structures leads to two final simplifications.\\[-2.4em]

\section{Conclusion and Future Work}
\label{sec:conclusion}
We have presented a method to construct algorithms to simplify expressions based on a set of equations (or axioms). We have shown that many parameters can change the efficiency and precision of such algorithms, and we have proposed a detailed experimental evaluation to show how they work. Our main conclusion is that the method can be useful in practice, but it must be carefully applied.

Many directions for future research are worth considering. We plan to apply our method to simplify various kinds of expressions, such as regular expressions (see \cite{stoughton2016formal}, for a  traditional approach) or usual algebraic expressions, for instance. To reach such goals, it will be useful to extend the method to more general sets of axioms such as Horn clauses \cite{DBLP:conf/lics/Kozen91,wechler1992universal}. Our current method naturally extends to such axiomatizations, since we already use the conditional application of axioms. The same kind of machinery can be used for triggering Horn clauses or, more generally, conditional axioms (i.e, inference rules).
We may also investigate using our method to simplify formulas in theorem provers.\\[-2em]

\section*{Acknowledgements}

\vspace{-0.5em}The author warmly thanks Pierre Flener, Gauthier van den Hove, and Jos\'e Vander Meulen for their useful comments about drafts of this paper.

\newpage
\label{sect:bib}
\bibliographystyle{plain}
\bibliography{easychair}

\setlength\textwidth{8.0in}

\newpage
\appendix
\section{Appendix: optional material for the reviewers}
\subsection{Goal of this appendix}
\label{app:goal}
This appendix is provided to help reviewers assess the contribution of this paper. More examples are given as well as more information about the expressions used in the experimental evaluation (Section \ref{sec:experiments}).
More information can be found at \\
{\footnotesize
 \hspace*{-0.0cm}\texttt{https://www.dropbox.com/sh/infjx6a9x7qc7qe/AABFygzGzWcTSIsSQNd0TdAOa?dl=0}.}
At this web address, the interested reviewers can also find the source code of the Java implementation of our algorithms as well as more test data. All program runs presented in this paper are executed on a MacBook Pro 2.7GHz (Intel
Core i5, 8Gb RAM) using Mac OS X 10.12.6. The programs are written in Java, and compiled
and executed using Java version \texttt{1.7.055}. Timings are measured using the method \texttt{System.nanoTime()}.

\subsection{Example expressions}

We unveil the expressions $\textrm{E}3$,  $\textrm{E}5$,  $\textrm{E}7$,  $\textrm{E}9$,  $\textrm{E}16$, $\textrm{B}3$, $\textrm{B}9$ used in Section \ref{sec:experiments}. Figure~\ref{fig:simplifiedexprs} shows the most simplified versions of the expressions while Figure~\ref{fig:bigexprs} provides the first five original expressions to be simplified; for $\textrm{B}3$ and $\textrm{B}9$, two files are available at the web address given at Section \ref{app:goal}.

\begin{figure}
\caption{Seven simplified expressions}
\ \\
\hspace*{-1.5cm}$\begin{array}{|l|c|}\hline
 \textrm{ES} 3 & \begin{minipage}{14.4cm}
\small
 \ \\[-0.1em]
\texttt{1}
 \ \\[-0.5em]
\end{minipage}
 \\
 \hline
 \textrm{ES} 5 & \begin{minipage}{14.4cm}
\small
 \ \\[-0.1em]
\texttt{(a + !b)(c + !(da))}
 \ \\[-0.5em]
\end{minipage}
 \\
 \hline
 \textrm{ES} 7 & \begin{minipage}{14.4cm}
\small
 \ \\[-0.1em]
\texttt{(e + c)(f(!b + g) + !(g + a + !b(e + d)))}
 \ \\[-0.5em]
\end{minipage}
 \\
 \hline
 \textrm{ES} 9 & \begin{minipage}{14.4cm}
\small
 \ \\[-0.1em]
\texttt{(c + d)(a + g) + !(ehf(ag + d)) + b + i }
 \ \\[-0.5em]
\end{minipage}
 \\
 \hline
 \textrm{ES} 16 & \begin{minipage}{14.4cm}
\small
 \ \\[-0.1em]
\texttt{m(a + g + !(n + d)(!b + j))(!(bl + d(!(g + l) + h))pk + do!(k + c + p)ibaf)}
 \ \\[-0.5em]
\end{minipage}
 \\
 \hline
 \textrm{BS} 3 & \begin{minipage}{14.4cm}
\small
 \ \\[-0.1em]
\texttt{!(a + b)c}
 \ \\[-0.5em]
\end{minipage}
 \\
 \hline
 \textrm{BS} 9 & \begin{minipage}{14.4cm}
\footnotesize
 \ \\[-0.1em]
\texttt{(f + (a + g)(e + b) + i)(!(c + d) + (d + !g(h + b))f!(a + (b!d + !(c + g)e)h(!e + i)))}
 \ \\[-0.5em]
\end{minipage}
 \\
 \hline
\end{array}$
\label{fig:simplifiedexprs}
\end{figure}

%

\begin{figure}
\vspace{-4cm}
\caption{Five big expressions}
\hspace*{-1cm}$\begin{array}{|l|c|}\hline
 \textrm{E} 3 & \begin{minipage}{13.0cm}
\footnotesize
 \ \\[-0.1em]
\texttt{a(!ac+!!b+a)(b+c+c(!b+c))+(a+!a+(a+b+a)!a+!aa(!a+b+a+a)+(b+a)(b+a)+(a+b+c!c+!(}
\texttt{c+b))(!a(!c+a)+!b+!b+!b+a))((!a+cc)c+!a)+(!a!!!(ba)+!(b(!a+a)(!c+c+!(!!b!!((b+}
\texttt{a+c!bca)(c+c+a))))))(!b(c+b+a)+c+!c!!(bc!aba!b))+!(!a(b+(a+cc)(c+caa)a))+((c+!}
\texttt{a)(c+b)+!b+b)(a+b)!(!!b+!!a+c)(a+b)(c+!a)(a+a+!c+c)+ab!!aa!((a(b+a)+c+cb+b+c)b}
\texttt{ba)(b+b)(!a+!c+!a)(!a+!(aa(a+!(bb))ac((c+c(!c+bc))a+(!a+c+b+a)!!b)))!((ab(c+!(}
\texttt{!c!c+a+b)!c(c+c))(a+b(c+c)+!b+cc)c(ac+a+!b!b+(b+b)b!c)ccb(b+b+a)(c(!a+ab)+!b((}
\texttt{b+c)a+bc)+!(!!a+c+ca+c))+!(bb!b)+c+!ca)!((b+!c)(!!(!cc+!a(c+!c)+!a)+!(ca+a)))+}
\texttt{!!(ac+!c+!((c(!b+a+c+(a+bc+a+c+!(!c+(a+c+b+!c+a)(!b+c+c+!b+!(ab)+c)))(a+a+c+ac}
\texttt{c+c+!c))+(a+c)!bc)!!(c+!c)c!a!(b+!c)+(cb+ba)(bc+a)+!(a+!c)(a+c+a)+!!!(c(ca!b(b}
\texttt{+a)+ca))+!caa+a))+(cb+a)!cc!(cba)(!!!!(b(!a+!c))+(!c+!a)cb))(!(!cbc)!c(ab+c+!b}
\texttt{)(!a+!b)!(c+a)(b+c+c)(a!a!b(b+b+a)+!b+!a+!c(ac+!c)b!b+bc)+!(!c+!(b+a!a+b+!!b+a}
\texttt{+!c+ab)))}
 \ \\[-0.5em]
\end{minipage}
 \\
 \hline
 \textrm{E} 5 & \begin{minipage}{13.0cm}
\footnotesize
 \ \\[-0.1em]
\texttt{(!(!!eb)+!(e+!ece)+!(!!aa)+!(bb)+!a+!be!b+(a+e)!c+!d+!(b!c)+!!!(!e(!(cbed)+ee!}
\texttt{b)+!(d+!(be)+bc!b+!bbc+b+a))+(!!(!dab)+!a+c+ca!(e+e))(d+b)(b+b(c+c+!d+b))(!!b(}
\texttt{e+ba)b+!!a+(c+!db(ca+!c)+!e!(!d+a))!d)+((!(ba!b+!b)+!a+!(cab)+!d+a)!b(c+c)c+!(}
\texttt{!((e+!a+ea+!c)(a+!e)+cb+(e+a+!c)(!b+a)+ec)+(!d+e)!d!c+d)((e+e)a!b+c+b+b)b!b(e+}
\texttt{!d)!(b!e)(!b+!b)+!((ba+a)!((c+!c+!c+(!d+e)(e+b+!(a+!a))+d+d+a+eb+b!bc)(!be+!b+}
\texttt{(a+c)!b+b+b)!cd))+!a!a+!!d+!((!bd+d+ac)!(aaa!c!ede)+(!(d!c+!a+ad+c)+(b+e+!e)(!}
\texttt{d+!(da)+a(e+c)))!(!b(ccd+!da)+!a+ce!a+a+!e+!(e!d))((c+e+c)(e+!a)+b)(deec+bc))!}
\texttt{(!b(e+!c)(a+d)))((c+a+dd(c+e)+!d)(a+b+d+!e!!(!e!cd))+e(d!b+!a+!dd+c)+!((!a+!a+}
\texttt{c!a)(!a+!b)+(!b+b)(ae+!c)))!(e+e+cb!b+!ddc)(c(!e+!b)+!b!a(!d+d+!(!d+ad)+!(a+!c}
\texttt{+!d))+!!d+!!b(!d+ab+!b+b+!d)a+!(a+b+c(b+bb)!!a)!!(!a+bb+a!(a+d)+e)+!a!a))!(!cd}
\texttt{(e+c+!e)(a!(dc)+!(!a+!a)!d))(!b+!(b+cd+a!d+!(!b+c+b)+(e+!aa)!d!c)+a)}
 \ \\[-0.5em]
\end{minipage}
 \\
 \hline
 \textrm{E} 7 & \begin{minipage}{13.0cm}
\footnotesize
 \ \\[-0.1em]
\texttt{(!!efe(c+g+dd+!b+!e)(e+(!((!b+e)df(!(g!de)+!(((!c+!a)(b+b)+d+g)g)))+g+!f+!(af)}
\texttt{+g+!c)(!a+b)+!((!f!d+!b+!a)(!e+(g+e)(g+e)+ec!d!!ec)b+f!d(!c+g+f)+ea+!d))+!g+!d}
\texttt{+!b+!c+c+e+e+dd+!a)!(a+g)(!gg+!g)(b+!((b+f+d+!!e+a+e)(!c+!g)+!!(ca))+!(!(a(c!!}
\texttt{c+!(!e!ac)+!f!cddaa+!e(!b+bg(a+d!dd+!f+!(!c!g)))+!d)!!(!cg+!!g+!(!!!!b(!g!b+c)}
\texttt{)!db+dg+b!e))+!(bf(ff+!g)+!(gb)+d+bb+e+b+f)!((g+!a+ceg+c)(!d+c)(f+f)e!(aa(bcf+}
\texttt{!d!f))+a!(!g+d)(!g+da+c+eg+!b))))(!!c+e)+(!b+g)(!a+!!a+!a)(!c(g+!d)+ef!!(!g!f)}
\texttt{+a+!a+g+b+d+df(d+f+f))!!(e+c+!f+e)(c+!d+!c+!!f+ea+d+g)f+((e+!f+(e+!gc+!bg+!a)f}
\texttt{c)cdf!!dd!ceeg+c!((c+f)!(d!a+!(e+a))))!(!(c+fg(f+a))+!a+b+!b+e+!d)((!c+!!!(fbg}
\texttt{c))(g+!e)+(g!d+f(!a+b+e)!f!f+d+!(b+(!g+!c)(d+a+c)))(!c+a+!b)(!f+c+f))(!(f+a+!f}
\texttt{+(g+!a+c)(!c+!a)!!b(!c+g+!g+a)+d+!(ae!f+ff))+!!g+!d+g!d)(dgb+!b)+!(!dad(!a+af)}
\texttt{+(c+g)!(ee(e+f))+g+g+!e+e+f(a+!d)+!a+e+!be+d+!g+g(!dg+!f+a+f))}
 \ \\[-0.5em]
\end{minipage}
 \\
 \hline
 \textrm{E} 9 & \begin{minipage}{13.0cm}
\footnotesize
 \ \\[-0.1em]
\texttt{d!!e(a+g)!a(!(!ffd)+e(b+b))(!!fb+(c!i+!!e)f)(!(a+g+!d!((b+!i)bga(bgi+h!fd))(gc}
\texttt{+e)+d+f+!!(d+h+h+!a)(!e(!g+!f)+!bh+!g!(e!e)+bgg))+hd(!g+hh)df(i+c+c)(g!i+d)(d+}
\texttt{i)+!(cc((!d+!e+b!c)(!d!h+f+(b+h)(i+h))+b!b!b(!i+!a)))!(a!e!a)(!f!c+a+!di)(!!h!}
\texttt{c(!ei+h+!b)!g+(ii+!c)(!(!!a+!a+!h)+!(!aa+e))h!d(!!hi+f!f+(i+e)(i+c)c)+!(!e!(((}
\texttt{!c+f+c)(!e+!e+b(h+f))+f+c+!e+!e)!!((g+c+b+!b)!ic)(b+ci+!i+c))))!(((i+!d)i+!(!c}
\texttt{ff))(!!f+ic!a)!!(f+(!(!eb)+de+a+!f)(!c+!g!i(a!c+c)(i+d(!g+b!d+(i+!c)!!e!!e))))}
\texttt{!!g((f+b)!g!g+!e(d+!g+!e))))hd!i(i+b!!b)((a+g)d!f+a)c!h+!!(!(b+e)+!f+(b+f+e)!(}
\texttt{c+h+g!g+g+!!(edc))+!(a+d)+!e+!!(!a!a(f+f+!d)g)+ag(i+g)h!b(i+!b!a)((!g+f+d)ai+g}
\texttt{c))+(!h!b+eb+!fb)(hb+!d+h+!c)+(!i+!!(((c+!d)!c!((a!!b+a)hf)!h+!h)(!i+f+(g(i+a)}
\texttt{+!(!bb))!f!(d+!e))))a(d+!h)+(f(!a+!i)(i+b)+!g+(i+(!!ba+c)!c)i+!(i+db)eca(e+e)!}
\texttt{b(d(!a+f+e)+(a+b)!(!eb!h)+g))ea+(eh(i+i+!d)+b+!h)(!ge+cfh+d+i)}
 \ \\[-0.5em]
\end{minipage}
 \\
 \hline
 \textrm{E} 16 & \begin{minipage}{13.0cm}
\footnotesize
 \ \\[-0.1em]
\texttt{(!(!m+!a)+(i!kde(e+!a)+(j+!b)!d)!n(!!oo+n(j+!!(l+c))(!!c+!kn)+!!dk(a+e)+!n)+g)}
\texttt{(!!((!k!o+!!m+p(i+m)!i)(!l+!b)p(k+!h)k!((p+h)!k))!(d!!!((!((g+o)e)+!(j!!!o)p+!}
\texttt{j+g+me)(g+lp)))+(!p+l+i)!!(b(!c+j+!d+k)(ab+!(!(f+!b)+f)))(k+!((ceji+!c+i+g+bm)}
\texttt{!!kdn+!((m+c+!d+!jm)(!p+!p)))+!!e(l!i+n)!m!(!k+n))(!((a+j)!n!k+!a)mg!g+d+!e+ep}
\texttt{)(p+f)!cdo!((!lg+m+p+!d)djcim(!e+i)m+k!(fh)+!i+hf(!a+p)+!(fbh)!(!!!(n!lp)+b))(}
\texttt{b+g))((!!((k(!h+ne)+!!a+f+h)(hd+!h+(!c!m+!a)(f+b!b))(b(j+ni)+a(f+n)+!(a(f+!o))}
\texttt{(f+i)!((!n+o)e)!(j+ol+((o+d+g+i)!!b!o+!j)!p))!((!h!n+!!(bp)(e+!g(fka+k)+b))(!(}
\texttt{!gb+bm)+a+!(!ei)+f+m!j+!((l+b)m+f)(b+d)+!(b+g+f+e)+p+!!c+!o)))+o+g+!g+g!h+(h+d}
\texttt{d)((!e!g+l+g)(o+l+!m)+(!jmf+e+n)(m+c)+a)+(!d+!(n+!g)(m+b)(b+!(c+!b)(h+!j+d)))!}
\texttt{d!l(i(m+!a+d)+h)+!(!p(!i+!h))+!o+h+i+f!n!c(b+d+o+pc+k)+ba+!g+!p)!d!h(!f+k+o)(d}
\texttt{+f+!k)!!(b+!e(!a+g+f)kp+!k+!((c+j)!k(d+!c)))(p!mg+!(!j+!(fd+!h)))+!h+!d+(!a+n+}
\texttt{!b+!b+m)!(k+!h))}
 \ \\[-0.5em]
\end{minipage}
 \\
 \hline
\end{array}$
\label{fig:bigexprs}
\end{figure}


 \subsection{Another example of how simplification takes place}
\label{app:method}

\begin{example}
\label{lab:ex1}

Let us simplify the boolean expression \texttt{a + ab}, using the axioms of Figure \ref{fig:axioms}. By hand, we can simplify it as follows:
\begin{center}\tt
\begin{tabular}{rccclccccrcccl}
 a + ab   &&=&& a1 + ab    &&&&&  $( x1$ && $=$ && $ x)$ \\
          &&=&& a(1 + b)    &&&&&  $( xy + xz$ && $=$ && $ x(y + z) )$ \\
          &&=&& a1    &&&&&  $( 1 + x$ && $=$ && $ 1 )$ \\
          &&=&& a    &&&&&  $( x1$ && $=$ && $ x )$ 
\end{tabular}
\end{center}
With our method, using valuations, the simplification works as follows: Remember that we generate valuations by enumerating identifiers in chronological order. We consider, in turn: $\id(\texttt{a})$, $\id(\texttt{b})$, $\id(\texttt{ab})$, $\id(\texttt{a + ab})$. The first identifier gives rise, among others, to the valuation $\{x \mapsto \id(\texttt{a})\}$, which is applied, among others, to the axiom $x\,1 = x$. This application creates the new structure
$.(\id(\texttt{a}), \id(\texttt{1})) : \id(\texttt{a})$. 
Later on, the identifier $\id(\texttt{b})$ is considered, the valuation $\{x \mapsto \id(\texttt{b})\}$ is generated, and it is applied to the axiom $1 + x = 1$, which creates the new structure
$+(\id(\texttt{1}), \id(\texttt{b})) : \id(\texttt{1})$.
Still later, the identifier $\id(\texttt{ab})$ is taken into account. At this step,
the valuation $\textit{val} \ = \ \{x \mapsto \id(\texttt{a}), y \mapsto \id(\texttt{1}), z \mapsto \id(\texttt{b})\}$ is generated because the two structures $.(\id(\texttt{a}), \id(\texttt{b})) : \id(\texttt{ab})$ and $+(\id(\texttt{a}), \id(\texttt{1})) : \id(\texttt{a})$ currently exist. 
This valuation is thus applied, among others, to the axiom $xy + xz\ =\ x(y + z) $. Therefore, the operation $\textit{toSet}$ is applied to $xy + xz$ and $\textit{val}$, which returns $\id(\texttt{a + ab})$ because $\textit{toSet}(xy, val)\ =\ \id(\texttt{a})$ (since $.(\id(\texttt{a}), \id(\texttt{1})) : \id(\texttt{a})$ exists) and because 
$\textit{toSet}(xz, val)\ =\ \id(\texttt{ab})$ (since $.(\id(\texttt{a}), \id(\texttt{b})) : \id(\texttt{ab})$ exists).
Symmetrically,  the operation $\textit{toSet}$ is applied to $x(y + z)$ and to $\textit{val}$, which returns $\id(\texttt{a})$ because the structures $.(\id(\texttt{a}), \id(\texttt{1})) : \id(\texttt{a})$ and $+(\id(\texttt{1}), \id(\texttt{b})) : \id(\texttt{1})$ exist.
Finally, the application of the axiom unifies the identifiers $\id(\texttt{a + ab})$ and $\id(\texttt{a})$, which renames $\id(\texttt{a + ab})$ into $\id(\texttt{a})$, everywhere in the collection of structures. This is the way \texttt{a + ab} is simplified into \texttt{a}, according to our method. But note also that the collection of structures now contains the structure $+(\id(\texttt{a}), \id(\texttt{ab})) : \id(\texttt{a})$. So the information that  \texttt{a + ab}\ =\ \texttt{a} is memorized.
\end{example}
\begin{note}


In practice, the algorithm implementing our method uses a global variable \texttt{mainId} that remains equal to the identifier $i$ of the sets of structures $E_i$ that contains the current minimal expression. The value of \texttt{mainId} is changed each time it is renamed. In the example above, we have, at the beginning, $\texttt{mainId} = \id(\texttt{a + ab})$, and after application of the axiom, we have $\texttt{mainId} = \id(\texttt{a})$.

In fact, after application of the axiom, we have  $\id(\texttt{a}) = \id(\texttt{a + ab})$ (for the current collection of structures). The algorithm may replace (the old value of) $\id(\texttt{a})$ by (the old value of) $\id(\texttt{a + ab})$, or conversely. But, in practice, we always rename the younger identifier into the older one. This is most efficient since it prevents the algorithm from computing valuations and applying axioms to a younger identifier that in fact is a renaming of an older one that has been previously processed. In the example above, the (old value of) identifier $\id(\texttt{a + ab})$ is removed from the collection of structures. So it is not taken into account at all for building valuations and applying them to axioms. The algorithm may terminate immediately after considering $\id(\texttt{ab})$.

The last axiom application presented in Example \ref{lab:ex1} uses a valuation in which the variable $y$ is mapped to the identifier $\id(1)$. Thus, this axiom application uses $1$ although the expression \texttt{ab}, thanks to which the valuation is generated, does not contain \texttt{1}. This shows that our method of computing valuations is more powerful than one can think at first glance. Note also that consideration of the expression \texttt{a + ab} would not have generated a valuation useful to simplify that expression itself.

\end{note}

\subsection{Difficult issues, solutions and workarounds}
\label{app:workaround}

This section is an expanded version of Section \ref{subsec:workaround}.

The algorithm of Figure \ref{fig:algo} is simple. 
However, experiments with this first version have revealed several weaknesses. Below, we address these problematic issues.

In the following, we call \emph{an iteration}, any execution of the body of the main loop of the algorithm, up to the garbage collector call. 
A major problem arises when an iteration is unable to consider each and all of the identifiers existing at the beginning of the iteration. (Looking at Figure \ref{fig:algo}, this means that the condition \textit{time}($i$) $>$ \texttt{timeLimit} is never evaluated to \emph{true} during a first execution of the body of the innermost loop of the algorithm (with $\texttt{reCount}=0$).)
In that case, some valuations 
are not generated; parts of the collection of structures remain unexplored. This leads to what we call \emph{an early-fixpoint}: the algorithm stops after simplifying some sub-expressions well but ignoring completely parts of the whole thing.

The above situation arises because too many valuations have been generated for the identifiers actually taken into account. So, we limit the number of generated valuations as follows. The major cause of the problem is a kind of unfairness: new structures are generated by applications of the axioms, and they can possibly be taken into account immediately for generating new valuations. Therefore, the algorithm has to consider many more valuations than those that would be generated at the start of the iteration. We solve this issue by considering only, for generating valuations, structures that were created before the current sub-iteration has started. (We call \emph{a sub-iteration}, any segment of the execution of the body of the innermost loop of the algorithm, between two successive evaluations to true of the condition \textit{time}($i$) $>$ \texttt{timeLimit} (see Figure \ref{fig:algo}).) Experiments have shown that this change 
is a big improvement in most cases.
However, after many other experiments, we have found it to be useful to add another limitation to valuation generation, as well as a similar limitation to axiom application: It may happen that an axiom application creates new structures that are unlikely to help in the simplification process because they are too large. So we limit the size of the identifiers in the valuations, and we limit the size of the structures created by axiom applications. The general rule is that a structure $f(i_1, i_2):i$ is acceptable for generating a valuation only if $\textit{size}(i_j) \leq \textit{size}(i)$ ($j=1,2$). Similarly, such a structure can be created only if  $\textit{size}(i_j) \leq \textit{size}(i')$ ($j=1,2$) where $i'$ is the existing identifier which is about to be unified with $i$. 

In many cases, the above changes 
actually solve the early-fixpoint problem and  make the algorithm more efficient without losing precision. 
But there are still situations where the proposed improvements are not powerful enough. Therefore, we have introduced additional workarounds that allow the algorithm to consider more identifiers, and to stop within predictable time. The first workaround consists of calling eraly a garbage collector inside an iteration whenever the memory assigned to the collection of structures becomes full before the condition \textit{time}($i$) $>$ \texttt{timeLimit} is first evaluated to true; we then continue  to iterate normally. The drawback of this expedient is that it can remove promising structures too early. 
Moreover, it is possible that, after some effective calls,  garbage collection recovers almost no memory or even no memory at all. In such cases, completing the iteration can take enormous time, without computing valuations for many identifiers. Thus, as a very last expedient, we add a time-out to each iteration. 


There is another early-fixpoint issue related to garbage collection: the garbage collector that is used initially by our algorithm keeps all minimal structures in the collection, i.e., all structures representing sub-expressions of one of the current minimal expressions. Let us call a \emph{plateau} a sequence of iterations during which the size of minimal expressions does not decrease. Using the standard garbage collector explained above, it is clear that minimal structures kept after some iterations of a plateau are necessarily also present after all subsequent iterations of the plateau. So, there is a big risk of getting a fixpoint collection with a possibly large number of structures. To deal with this problem, we use other forms of garbage collectors, the most restrictive of which only keeps structures representing a single currently minimal expression. The most recent structures are kept.
Then, each time we enter a plateau, we switch to another garbage collector, but we stick to the same when the current size of minimal expressions decreases. The intuition is that, when we are stuck in a plateau, we must ``shake'' the collection of structures to open a new avenue where simplification is possible anew.

\subsection{Execution of other variants of the algorithm}
\begin{table}[t!]
\caption{Execution of algorithm \textrm{var/12} on expression \textrm{E}9}
\vspace{-0.5cm}
$$\begin{array}{|r|r||r|r|r||r|r|r|r||r| r||r|r|r|r|}\hline
\textrm{time}  & \textrm{size} & \textrm{nval} &  \textrm{napl} & \textrm{a} / \textrm{v} & \textrm{ds01}  & \textrm{nd01} & \textrm{ods} & \textrm{nods} & \textrm{nid}_1 & \textrm{nid}  & M_i & N_i & M_f & N_f \\\hline\hline
12.94 & 174 & 3.6M & 538K & 0.1 & 508 & 45 & 118 & 55 & 333 & 282 & 373 & 372 & 42K & 540K\\
3.03 & 160 & 749K & 306K & 0.4 & 7 & 3 & 7 & 4 & 164 & 163 & 164 & 315 & 29K & 325K\\
23.57 & 76 & 3.4M & 4.7M & 1.3 & 59 & 11 & 25 & 12 & 2.9K & 505 & 359 & 854 & 38K & 2.5M\\
11.01 & 36 & 1.2M & 2.9M & 2.4 & 27 & 4 & 13 & 7 & 4.0K & 264 & 87 & 205 & 31K & 2.5M\\
\hline
10.07 & 35 & 972K & 2.9M & 3.0 & 0 & 0 & 1 & 1 & 2.5K & 240 & 67 & 203 & 21K & 2.5M\\
10.97 & 31 & 1.1M & 3.0M & 2.7 & 0 & 0 & 4 & 1 & 4.9K & 265 & 98 & 394 & 20K & 2.5M\\
13.54 & 31 & 1.3M & 3.2M & 2.3 & 0 & 0 & 0 & 0 & 2.1K & 255 & 58 & 261 & 21K & 2.5M\\
11.74 & 27 & 1.2M & 3.0M & 2.5 & 0 & 0 & 4 & 2 & 7.1K & 270 & 87 & 218 & 21K & 2.5M\\
11.45 & 25 & 1.1M & 2.9M & 2.5 & 0 & 0 & 2 & 1 & 5.2K & 263 & 81 & 229 & 23K & 2.5M\\
11.82 & 25 & 1.1M & 3.0M & 2.6 & 0 & 0 & 0 & 0 & 3.1K & 256 & 74 & 180 & 24K & 2.5M\\
0.01 & 24 & 4.4K & 4.0K & 0.9 & 0 & 0 & 1 & 1 & 243 & 41 & 25 & 24 & 435 & 4.2K\\
\hline
\end{array}$$
\label{tab:labd}
\end{table}

Let us examine the execution of a version of the algorithm very different from the default, namely $\textrm{var/}12$. Table \ref{tab:labd} provides the relevant information. Note that this variant is the most precise and fastest of the $32$ versions that use valuations of type $3$. This version uses early-application of axioms involving a single variable (and, in fact, the $15$ fastest versions using this kind of valuations also use early-application of these axioms). We see that the first four iterations do most of the job but they take much more time than the iterations of the default version (roughly  more than ten times). Many more valuations are generated but they are less productive (which was expected). In this table, column $\textrm{nid}_1$ gives the number of times all axioms involving a single variable are applied to an identifer, while 
$\textrm{nid}$ is the number of times other axioms are applied to the valuations generated for a single identifier. We see that the values in $\textrm{nid}_1$ are much greater than the corresponding values in $\textrm{nid}$, which means that a big part of the work done by early-application is not exploited by the other axioms. The right column of the table ($N_f$) shows that almost all iterations stop because no room is left for creating new structures. This version of the algorithm really uses a kind of breadth-first strategy while the default version uses a more balanced one. In fact, with this version of the algorithm, there is a big risk of getting an early-fixpoint, although it is not actually the case for expression $\textrm{E} 9$. But the problem arises for expression $\textrm{B} 9$, as shown in Table \ref{lab:synthesisSuperTests}.

\begin{table}[t!]
\caption{Execution of algorithm \textrm{var/4} on expression \textrm{B}9}
\vspace{-0.5cm}
$$\begin{array}{|r|r||r|r|r||r|r|r|r||r||r|r|r|r|}\hline
\textrm{time}  & \textrm{size} & \textrm{nval} &  \textrm{napl} & \textrm{a} / \textrm{v} & \textrm{ds01}  & \textrm{nd01} & \textrm{ods} & \textrm{nods} & \textrm{nid}  & M_i & N_i & M_f & N_f \\\hline\hline
2.94 & 15445 & 106K & 496K & 4.6 & 71071 & 2636 & 13484 & 5041 & 22K & 33K & 33K & 72K & 540K\\
0.57 & 11149 & 38K & 221K & 5.8 & 3565 & 102 & 731 & 205 & 9.0K & 9.3K & 23K & 67K & 289K\\
1.1 & 8850 & 64K & 265K & 4.1 & 1620 & 66 & 679 & 87 & 10K & 17K & 45K & 26K & 171K\\
0.31 & 8571 & 20K & 121K & 6.1 & 154 & 20 & 125 & 37 & 5.2K & 5.3K & 11K & 33K & 155K\\
1.23 & 7419 & 63K & 269K & 4.2 & 373 & 53 & 779 & 54 & 11K & 11K & 32K & 19K & 147K\\
0.17 & 6359 & 11K & 72K & 6.8 & 974 & 2 & 86 & 15 & 3.6K & 3.6K & 7.0K & 17K & 91K\\
0.51 & 6216 & 29K & 151K & 5.2 & 61 & 11 & 82 & 22 & 5.2K & 5.8K & 15K & 48K & 194K\\
1.12 & 4808 & 62K & 235K & 3.8 & 394 & 43 & 1014 & 22 & 8.9K & 14K & 41K & 17K & 131K\\
0.96 & 4398 & 62K & 260K & 4.1 & 336 & 39 & 74 & 29 & 10K & 2.5K & 5.0K & 65K & 290K\\
0.64 & 4215 & 46K & 201K & 4.3 & 27 & 10 & 156 & 17 & 5.8K & 6.6K & 21K & 61K & 249K\\
1.8 & 3694 & 106K & 383K & 3.6 & 301 & 41 & 220 & 16 & 15K & 12K & 46K & 36K & 267K\\
1.02 & 3187 & 63K & 256K & 4.0 & 445 & 18 & 62 & 34 & 8.7K & 2.2K & 4.9K & 67K & 286K\\
0.37 & 3133 & 28K & 125K & 4.5 & 5 & 3 & 49 & 8 & 3.9K & 4.3K & 14K & 39K & 160K\\
1.22 & 3063 & 77K & 268K & 3.4 & 51 & 13 & 19 & 6 & 10K & 8.3K & 32K & 20K & 152K\\
0.6 & 2845 & 41K & 170K & 4.1 & 160 & 8 & 58 & 15 & 6.3K & 1.6K & 3.1K & 44K & 194K\\
0.51 & 2808 & 35K & 146K & 4.1 & 8 & 4 & 29 & 4 & 3.8K & 4.4K & 14K & 42K & 173K\\
1.6 & 2755 & 96K & 322K & 3.3 & 15 & 2 & 38 & 10 & 11K & 11K & 45K & 32K & 220K\\
0.81 & 2639 & 52K & 208K & 3.9 & 97 & 12 & 19 & 9 & 6.5K & 1.7K & 3.7K & 54K & 231K\\
0.52 & 2519 & 35K & 145K & 4.1 & 19 & 6 & 101 & 6 & 3.5K & 4.2K & 15K & 38K & 163K\\
2.1 & 1795 & 124K & 422K & 3.4 & 675 & 14 & 49 & 12 & 16K & 7.2K & 28K & 28K & 268K\\
1.74 & 128 & 93K & 332K & 3.5 & 1635 & 4 & 32 & 10 & 12K & 930 & 1.7K & 75K & 370K\\
0.67 & 111 & 39K & 160K & 4.1 & 9 & 3 & 8 & 3 & 4.2K & 108 & 234 & 25K & 152K\\
2.42 & 88 & 126K & 494K & 3.9 & 0 & 0 & 23 & 12 & 7.0K & 146 & 439 & 75K & 465K\\
1.28 & 79 & 71K & 288K & 4.0 & 3 & 2 & 6 & 5 & 6.0K & 131 & 467 & 42K & 259K\\
0.46 & 76 & 28K & 121K & 4.3 & 0 & 0 & 3 & 2 & 2.9K & 84 & 187 & 20K & 119K\\
1.55 & 72 & 79K & 350K & 4.4 & 4 & 1 & 0 & 0 & 6.7K & 66 & 125 & 53K & 319K\\
0.58 & 69 & 32K & 140K & 4.3 & 1 & 1 & 2 & 2 & 3.4K & 81 & 192 & 25K & 135K\\
0.8 & 54 & 44K & 193K & 4.3 & 8 & 2 & 7 & 4 & 4.0K & 88 & 210 & 34K & 184K\\
\hline
1.89 & 52 & 102K & 427K & 4.1 & 0 & 0 & 2 & 1 & 5.6K & 57 & 117 & 75K & 407K\\
0.83 & 52 & 44K & 199K & 4.4 & 0 & 0 & 0 & 0 & 4.0K & 70 & 181 & 33K & 185K\\
2.08 & 50 & 96K & 397K & 4.1 & 0 & 0 & 2 & 2 & 5.8K & 420 & 1.5K & 75K & 382K\\
\hline
\end{array}$$
\label{tab:labe}
\end{table}

Finally, Table \ref{tab:labe} shows how well the algorithm \textrm{var/4} behaves for the very large expression $\textrm{B} 9$. There is only one short plateau when the size is already reduced to $52$. Remember that the four non standard axioms in the bottom right corner of Figure \ref{fig:axioms} are used by this variant. It can be seen that the simplification power of the algorithm does not diminish after a few iterations: at some iterations, it just ``massages'' the collection of structures to find better simplification opportunities; then it starts again with new axiom applications that use $0$ or $1$, until a very small expression is obtained.




\end{document}